\documentclass[a4paper,10pt]{article}
\usepackage[dvips]{graphicx}
\usepackage{amssymb,amsmath}
\numberwithin{equation}{section}

\begin{document}

 \def\gsim{ \lower .75ex \hbox{$\sim$} \llap{\raise .27ex \hbox{$>$}} }
 \def\lsim{ \lower .75ex \hbox{$\sim$} \llap{\raise .27ex \hbox{$<$}} }



 \title{ The  (2+1)-dimensional Hirota-Maxwell-Bloch equation: Darboux transformation and soliton solutions}

\author{ Kuralay Yesmakhanova, Gaukhar Shaikhova, Kuanysh Zhussupbekov \\ and Ratbay  Myrzakulov\footnote{The corresponding author.
Email: rmyrzakulov@gmail.com}
 \\ \textit{Eurasian International Center for Theoretical Physics and  Department of General } \\ \textit{ $\&$  Theoretical Physics, Eurasian National University, Astana 010008, Kazakhstan}}

\date{}
 \maketitle


 \renewcommand{\baselinestretch}{1.1}

 \begin{abstract}
The (2+1)-dimensional Hirota-Maxwell-Bloch equation (HMBE)  is integrable by the Inverse Scattering Method. In this paper, we construct a Darboux transformation (DT) of the (2+1)-dimensional HMBE. Also the one-soliton solution  obtained by means of the  one-fold DT. For the $n$-soliton solution the general form is presented. 
 \end{abstract}

 
\section{Introduction}
Modern nonlinear science as a powerful subject explains all kinds of mysteries in the challenges of modern  technology and science. The nonlinear nature of the real systems is considered to be fundamental to the understanding of most natural phenomena. Integrable systems are the main part of theory of modern nonlinear science. One of the interesting integrable system is the so-called (1+1)-dimensional  Hirota -Maxwell-Bloch equation. It describes the nonlinear dynamics of femtosecond pulse propagation through doped fibre.  In this paper, we consider one of  (2+1)-dimensional integrable generalizations of the (1+1)-dimensional  HMBE, namely, the (2+1)-dimensional  HMBE. We present the Darboux transformation and using it,  the one-soliton solution.

The paper is organized as follows.  In Section 2,  we give a brief review of the (1+1)-dimensional  HMBE.  The  (2+1)-dimensional HMBE we present in Section 3. In Section 4, we construct the DT for  the  (2+1)-dimensional HMBE.  In  Section 5, using the constructed  one-fold DT, the one-soliton  solution of the  (2+1)-dimensional HMBE is given. In Section 6,  conclusions are given.

\section{Brief review of the (1+1)-dimensional  Hirota -Maxwell-Bloch equation}
To establish our notation, definitions and terminoloy let us first recall some main informations on the (1+1)-dimensional Hirota -Maxwell-Bloch equation (HMBE). The (1+1)-dimensional HMBE  has the form (see e.g. \cite{C1}-\cite{C3}, \cite{R13}-\cite{R17})
 \begin{eqnarray}
iq_{t}+\epsilon_1(q_{xx}+2\delta|q|^2q)+i\epsilon_2(q_{xxx}+6\delta|q|^2q_x)-2ip&=&0, \label{2.1}\\
p_{x}-2i\omega p -2\eta q&=&0,\label{2.2}\\
\eta_{x}+\delta(q^{*} p +p^{*} q)&=&0,\label{2.3}
 \end{eqnarray}
  where  $q, p$ are complex functions, $\eta$ is a real function and $\omega, \epsilon_i$ are real constants. By setting $\delta=+1$  or $\delta=-1$, the HMBE with attractive or repulsive interaction is obtained. The functions $p$ and $\eta$ satisfy the relation
   \begin{eqnarray}
\eta^{2}+\delta|p|^{2}=b(t)=cons(t).\label{2.4}
 \end{eqnarray}
  The DT and some exact solutions of this equation were presented in \cite{C1}-\cite{C3}. Note that the gauge \cite{R13} and Lakshmanan equivalent counterpart \cite{R16} of the (1+1)-dimensional HMBE (\ref{2.1})-(\ref{2.3})  is the following  Myrzakulov-XCIV equation  (M-XCIV equation) \cite{R13}: 
\begin{eqnarray}
iS_{t}+\frac{1}{2}\epsilon_1[S,S_{xx}]+i\epsilon_2(S_{xx}+6tr(S_x^2)S)_{x}+\frac{1}{\omega}[S, W]&=&0,\label{2.5}\\
 iW_{x}+\omega [S, W]&=&0,\label{2.6}
\end{eqnarray} 
where $S=S_{i}\sigma_{i}$, $W=W_{i}\sigma_{i}$, $S^{2}=I, \quad W^{2}=b(t)I$, $b(t)=const(t)$, $I=diag(1,1)$,  $[A,B]=AB-BA$, $\omega$ is a real constant and $\sigma_{i}$ are Pauli matrices. The DT of the M-XCIV equation and its one-soliton solution were constructed in \cite{R17}.
  
\section{The (2+1)-dimensional  Hirota -Maxwell-Bloch equation}
 
  The (2+1)-dimensional  Hirota -Maxwell-Bloch equation (HMBE) reads as \cite{R14}
 \begin{eqnarray}
iq_{t}+\epsilon_1q_{xy}+i\epsilon_2q_{xxy}-vq+i(wq)_x-2ip&=&0, \label{3.1}\\
v_{x}+2\epsilon_1\delta(|q|^2)_y-2i\epsilon_2\delta(q^{*}_{xy}q-q^{*}q_{xy})&=&0,\label{3.2}\\
w_{x}-2\epsilon_2\delta(|q|^2)_y&=&0,\label{3.3}\\
p_{x}-2i\omega p -2\eta q&=&0,\label{3.4}\\
\eta_{x}+\delta(q^{*} p +p^{*} q)&=&0,\label{3.5}
 \end{eqnarray}
 where  $q, p$ are complex functions, $v,w, \eta$ are real functions.   This  set of equations  (\ref{3.1})-(\ref{3.5})  is  integrable by IST. 
 The corresponding Lax representation reads as
\begin{eqnarray}
\Psi_{x}&=&A\Psi,\label{3.6}\\
\Psi_{t}&=&(2\epsilon_1\lambda+4\epsilon_2\lambda^2)\Psi_y+B\Psi,\label{3.7} 
\end{eqnarray}  
where  
 \begin{eqnarray}
A&=&-i\lambda \sigma_3+A_0,\label{3.8}\\
B&=&\lambda B_1+B_0+\frac{i}{\lambda+\omega}B_{-1}.\label{3.9} 
\end{eqnarray} 
Here
\begin{eqnarray}
B_1&=&iw\sigma_3+2i\epsilon_2\sigma_3A_{0y},\label{3.10}\\
A_0&=&\begin{pmatrix} 0&q\\-r& 0\end{pmatrix},\label{3.11}\\
B_0&=&-\frac{i}{2}v\sigma_3+\begin{pmatrix} 0&i\epsilon_1q_y-\epsilon_2q_{xy}-wq\\i\epsilon_1r_y+\epsilon_2r_{xy}+wr& 0\end{pmatrix},\label{3.12}\\
B_{-1}&=&\begin{pmatrix} \eta&-p\\-k& -\eta\end{pmatrix}\label{3.13} 
\end{eqnarray}
and $r=\delta q^{*}, \quad k=\delta p^{*}$, where $\delta=\pm1$. 
 The spectral parameter  $\lambda$ evolves as 
\begin{eqnarray}
\lambda_{t}=(2\epsilon_1 \lambda+4\epsilon_2 \lambda^2)\lambda_y.\label{3.14} 
 \end{eqnarray} 
 In this paper we restrict ourselves  to the case $\delta=+1$ that is to  the focusing (attractive interaction) case. We note that if $y=x$ the system  (\ref{3.1})-(\ref{3.5}) reduces to the (1+1)-dimensional HMBE (\ref{2.1})-(\ref{2.3}). This fact is explains why we called the system  (\ref{3.1})-(\ref{3.5}) as the (2+1)-dimensional HMBE. At last, we present the Myrzakulov-Lakshmanan-IV equation (ML-IV equation) \cite{R14}
\begin{eqnarray}
iS_{t}+2\epsilon_1Z_x+i\epsilon_2(S_{xy}+[S_x,Z])_{x}+(fS)_{x}+\frac{1}{\omega}[S, W]&=&0,\label{3.15}\\
u_x-\frac{i}{4}tr(S\cdot[S_x,S_y])&=&0,\label{3.16}\\
f_x-\frac{i}{4}\epsilon_2[tr(S_x^2)]_y&=&0,\label{3.17}\\
 iW_{x}+\omega [S, W]&=&0,\label{3.18}
\end{eqnarray}
where 
\begin{eqnarray}
Z=0.5([S,S_{y}+2iuS).\label{3.19}
\end{eqnarray}
As it was shown in \cite{R14}, the ML-IV equation is the equivalent counterpart of the (2+1)-dimensional HMBE. In 1+1 dimensions, the ML-IV equation reduces to the M-XCIV equation (\ref{2.5})-(\ref{2.6}).
 \section{DT for the (2+1)-dimensional HMBE}

In this section we construct the DT for  the (2+1)-dimensional HMBE (\ref{3.14})-(\ref{3.17}). In particular, we give in detail the one-fold DT and briefly the $n$-fold DT. 

\subsection{One-fold DT}
Let $\Psi$ and $\Psi^{\prime}$ are two solutions of the system (\ref{3.6})-(\ref{3.7}) so that
\begin{eqnarray}
\Psi'_{x} &=& A'\Psi',\label{4.1}\\
\Psi'_{t} &=&(2\epsilon_1\lambda+4\epsilon_2\lambda^2)\Psi'_{y}+B'\Psi'. \label{4.2}
\end{eqnarray}
We assume that these two solutions are related by the following transformation:
\begin{eqnarray}
\Psi'=T\Psi=(\lambda I-M)\Psi. \label{4.3}
\end{eqnarray}
The  matrix function $T$ obeys the following equations
\begin{eqnarray}
T_{x}+TA &=& A'T,\label{4.4} \\
T_{t}+TB &=&(2\epsilon_1\lambda+4\epsilon_2\lambda^2)T_{y}+B'T. \label{4.5}
\end{eqnarray}
From the equation (4.4) we get
   \begin{eqnarray}
		\lambda^0&:&  M_{x}=A'_{0}M-MA_0, \label{4.6}\\
	 \lambda^1&:& A'_{0}=A_{0}+ i[M,\sigma_{3}], \label{4.7}\\
 \lambda^2&:& iI\sigma_{3}=i\sigma_{3}I. \label{4.8}
\end{eqnarray}
Eq.(\ref{4.7}) gives 
\begin{eqnarray}
q^{[1]} &=& q-2im_{12}, \label{4.9}\\
q^{*[1]} &=&  q^{*}-2im_{21},  \label{4.10}
\end{eqnarray}
where  
\begin{eqnarray}
M=\begin{pmatrix} m_{11}&m_{12}\\m_{21}&m_{22}\end{pmatrix},\quad I=\begin{pmatrix} 1&0\\ 0&1\end{pmatrix}. \label{4.11}
\end{eqnarray}Hence we get $m_{21}=- m_{12}^{*}$ in our    attractive interaction case. Eq.(\ref{4.5}) gives us the following relations
\begin{eqnarray} 
I&:&\lambda_{t}=(2\epsilon_1 \lambda+4\epsilon_2 \lambda^2)\lambda_y,\label{4.12}\\
 \lambda^{0}&:& -M_{t}=iB'_{-1}-B'_{0}M-iB_{-1}+MB_{0}, \label{4.13}\\
\lambda^{1}&:& 2\epsilon_{1}M_{y}=B'_{0}-B_{0}+MB_{1}-B_{1}^{'}M, \label{4.14}\\
\lambda^{2}&:& 4\epsilon_{2}M_{y}=B'_{1}-B_{1}, \label{4.15}\\
(\lambda+\omega)^{-1}&:&0=-i\omega B'_{-1}-iB'_{-1}M+i\omega B_{-1}+iMB_{-1}. \label{4.16}
\end{eqnarray}
Hence we get the DT
\begin{eqnarray} 
 B'_{0}&=&B_{0}-MB_{1}+(B_{1}+4\epsilon_{2}M_{y})M+2\epsilon_{1}M_{y}, \label{4.17}\\
B'_{1}&=&B_{1}+4\epsilon_{2}M_{y}, \label{4.18}\\
B'_{-1}&=&(M+\omega I)B_{-1}(M+\omega I)^{-1}. \label{4.19}
\end{eqnarray}
At the same, from Eqs.(\ref{4.17})-(\ref{4.18}) we get 
\begin{eqnarray}
v'&=&v+4i\epsilon_{1}m_{11y}+4m_{11}w+4\epsilon_{2}(m_{12}q^{*}_{y}-m_{12}^{*}q_{y}+2im_{11}m_{11y}-2im_{12}^{*}m_{12y}), \label{4.20}\\
w'&=&w-4i\epsilon_{2}m_{11y}=w+4i\epsilon_{2}m_{22y} \label{4.21}
\end{eqnarray}
and  we additionally  have  $m_{22}=m_{11}^{*}$. So the matix $M$ has the form
\begin{equation}
M=\begin{pmatrix} m_{11} & m_{12}\\ -m_{12}^{*} & m_{11}^{*}\end{pmatrix}, \quad M^{-1}=\frac{1}{|m_{11}|^2+|m_{12}|^{2}}\begin{pmatrix} m_{11}^{*} & -m_{12}\\ m_{12}^{*} & m_{11}\end{pmatrix},  \label{4.22}
\end{equation}
\begin{equation}
M+\omega I=\begin{pmatrix} m_{11}+\omega & m_{12}\\ -m_{12}^{*} & \omega+m_{11}^{*}\end{pmatrix},\quad 
(M+\omega I)^{-1}=\frac{1}{\square}\begin{pmatrix} m_{11}^{*}+\omega & -m_{12}\\ m_{12}^{*} & \omega+m_{11}\end{pmatrix}, \label{4.23}
\end{equation}
where
\begin{equation}
\square=det(M+\omega I)=\omega^2+\omega(m_{11}+m_{11}^{*})+|m_{11}|^{2}+|m_{12}|^{2}. \label{4.24}
\end{equation}
The equation (\ref{4.19}) gives  
\begin{eqnarray}
\eta'&=& \frac{(|\omega+m_{11}|^{2}-|m_{12}|^{2})\eta-pm_{12}^{*}(\omega+m_{11})- p^{*}m_{12}(\omega+m_{11}^{*})}{\square}, \label{4.25}\\
p'&=& \frac{p(\omega+m_{11})^{2}-p^{*}m_{12}^{2}+2\eta m_{12}(\omega+m_{11})}{\square},\\ \label{4.26}
{p^{*}}'&=&\frac{p^{*}(\omega+m_{11}^{*})^{2}-pm_{12}^{*2}
+2\eta m_{12}^{*}(\omega+m_{11}^{*})}{\square}. \label{4.27}
\end{eqnarray}
 We now assume that 
\begin{eqnarray}
M=H\Lambda H^{-1}, \label{4.28}
\end{eqnarray}
where \begin{eqnarray}
H=\begin{pmatrix} \psi_{1}(\lambda_{1};t,x,y)&\psi_{1}(\lambda_{2};t,x,y)\\\psi_{2}(\lambda_{1};t,x,y)&\psi_{2}(\lambda_{2};t,x,y)\end{pmatrix}. \label{4.29}
\end{eqnarray}
Here
\begin{eqnarray}
\Lambda&=&\begin{pmatrix} \lambda_{1}&0\\0&\lambda_{2}\end{pmatrix} \label{4.30}
\end{eqnarray}
and $det$ $H\neq0$, where $\lambda_{1}$ and $\lambda_2$ are complex constants.
The matrix $H$ obeys the system
\begin{eqnarray}
H_{x} &=& -i\sigma_{3}H\Lambda+A_{0}H, \label{4.31}\\
H_{t} &=& 2 H_{y}\Lambda +B_{0}H+B_{-1}H\Sigma, \label{4.32}
\end{eqnarray}
where 
\begin{eqnarray}
\Sigma=\begin{pmatrix} \frac{i}{\lambda_{1}+\omega}&0\\0&\frac{i}{\lambda_{2}+\omega}\end{pmatrix}. \label{4.33}
\end{eqnarray}
  In order to satisfy the constraints of $S$ and  $B'_{-1}$ as mentioned above, we first notes that 
\begin{eqnarray}
\Psi^{+}=\Psi^{-1}, \quad A_{0}^{+}=-A_{0}, \label{4.34}
\end{eqnarray} 
\begin{eqnarray}
\lambda_{2}=\lambda^{*}_{1}, \quad
 H=\begin{pmatrix} \psi_{1}(\lambda_{1};t,x,y)&-\psi^{*}_{2}(\lambda_{1};t,x,y)\\ \psi_{2}(\lambda_{1};t,x,y)&\psi^{*}_{1}(\lambda_{1};t,x,y)\\ \end{pmatrix}, \label{4.35}
\end{eqnarray}
\begin{eqnarray}
H^{-1}=\frac{1}{\Delta}\begin{pmatrix} \psi^{*}_{1}(\lambda_{1};t,x,y)&\psi^{*}_{2}(\lambda_{1};t,x,y)\\ -\psi_{2}(\lambda_{1};t,x,y)&\psi_{1}(\lambda_{1};t,x,y)\\ \end{pmatrix},\label{4.36}
\end{eqnarray}
where 
\begin{eqnarray}
\Delta &=&|\psi_{1}|^2+|\psi_{2}|^2. \label{4.37}
\end{eqnarray}
So  the matrix $M$ has the form
\begin{eqnarray}
M&=&\frac{1}{\Delta}\begin{pmatrix} \lambda_{1}|\psi_{1}|^2+\lambda_{2}|\psi_{2}|^2 & (\lambda_{1}-\lambda_{2})\psi_{1}\psi_{2}^{*}\\ (\lambda_{1}-\lambda_{2})\psi_{1}^{*}\psi_{2} & \lambda_{1}|\psi_{2}|^2+\lambda_{2}|\psi_{1}|^2)\end{pmatrix}.  \label{4.38}
\end{eqnarray}
Finally we can write  the one-fold  DT for  the (2+1)-dimensional HMBE as:
\begin{eqnarray}
q^{[1]}& = &q-2im_{12}, \label{4.39}\\
v^{[1]}&=&v+4i\epsilon_{1}m_{11y}+4m_{11}w+4\epsilon_{2}(m_{12}q^{*}_{y}-m_{12}^{*}q_{y}-2im_{11}m_{11y}+4im_{12}^{*}m_{12y}), \label{4.40}\\
w^{[1]}&=&w-4i\epsilon_{2}m_{11y}=w+4i\epsilon_{2}m_{22y}, \label{4.41}\\
\eta^{[1]}&=& \frac{(|\omega+m_{11}|^{2}-|m_{12}|^{2})\eta-pm_{12}^{*}(\omega+m_{11})- p^{*}m_{12}(\omega+m_{11}^{*})}{\square}, \label{4.42}\\
p^{[1]}&=& \frac{p(\omega+m_{11})^{2}-p^{*}m_{12}^{2}+2\eta m_{12}(\omega+m_{11})}{\square}. \label{4.43}
\end{eqnarray}
At last, we note that the expressions of $m_{ij}$ can be   rewritten  in the determinant form as 
\begin{eqnarray}
m_{11}=\frac{\lambda_{1}|\psi_{1}|^2+\lambda_{2}|\psi_{2}|^2}{\Delta}=\frac{\Delta_{11}}{\Delta}, \quad 
m_{12}=\frac{(\lambda_{1}-\lambda_{2})\psi_{1}\psi_{2}^{*}}{\Delta}=\frac{\Delta_{12}}{\Delta}, \label{4.44}
\end{eqnarray}
where
\begin{eqnarray}
\Delta_{11}=det\begin{pmatrix} \psi_{1} & -\lambda_{2}\psi_{2}^{*}\\ 
\psi_{2} & \lambda_{1}\psi_{1}^{*}\end{pmatrix}, \quad \Delta_{12}=-det\begin{pmatrix} \psi_{1} & \lambda_{1}\psi_{1}\\ 
\psi_{2}^{*} & \lambda_{2}\psi_{2}^{*}\end{pmatrix}. \label{4.45}
\end{eqnarray}

\subsection{$n$-fold DT}

In the previous subsection we have constructed the one-fold DT. Similarly we can construct the $n$-fold DT. To construct  the $n$-fold DT  of the (2+1)-dimensional HMBE, we note that the function $T$ satisfies the following equations
\begin{eqnarray}
T_{nx}&=&A^{[n]}T_n-T_{n}A, \label{4.46}\\
T_{nt}&=&(2\epsilon_{1}\lambda+4\epsilon_{2}\lambda^{2}) T_{ny}+B^{[n]}T_{n}-T_{n}B. \label{4.47}
\end{eqnarray}
The solution of this system can be written as \cite{C1}
\begin{eqnarray}T_n(\lambda;\lambda_1,\lambda_2,\lambda_3,\lambda_4,\dots,\lambda_{2n})=\lambda^n I+t_{n-1}^{[n]}\lambda^{n-1}+\dots+t_1^{[n]}\lambda+t_0^{[n]}=\frac{1}{\Delta_n}\left(\begin{matrix}T_{n11}&T_{n12}\\
T_{n21}&T_{n22}
\end{matrix}\right), \label{4.48}
\end{eqnarray}
where  \cite{C1}
\begin{eqnarray} \Delta_n&=&
\tiny\left|\begin{matrix}\Phi_{1,1}&\Phi_{2,1}&\lambda_1\Phi_{1,1}&\lambda_1\Phi_{2,1}&\dots&\lambda_1^{n-1}\Phi_{1,1}&\lambda_1^{n-1}\Phi_{2,1}\\
\Phi_{1,2}&\Phi_{2,2}&\lambda_2\Phi_{1,2}&\lambda_2\Phi_{2,2}&\dots&\lambda_2^{n-1}\Phi_{1,2}&\lambda_2^{n-1}\Phi_{2,2}\\
\Phi_{1,3}&\Phi_{2,3}&\lambda_3\Phi_{1,3}&\lambda_3\Phi_{2,3}&\dots&\lambda_3^{n-1}\Phi_{1,3}&\lambda_3^{n-1}\Phi_{2,3}\\
\Phi_{1,4}&\Phi_{2,4}&\lambda_4\Phi_{1,4}&\lambda_4\Phi_{2,4}&\dots&\lambda_4^{n-1}\Phi_{1,4}&\lambda_4^{n-1}\Phi_{2,4}\\
\vdots&\vdots&\vdots&\vdots&\vdots&\vdots&\vdots\\
\Phi_{1,2n-1}&\Phi_{2,2n-1}&\lambda_{2n-1}\Phi_{1,2n-1}&\lambda_{2n-1}\Phi_{2,2n-1}&\dots&\lambda_{2n-1}^{n-1}\Phi_{1,2n-1}&\lambda_{2n-1}^{n-1}\Phi_{2,2n-1}\\
\Phi_{1,2n}&\Phi_{2,2n}&\lambda_{2n}\Phi_{1,2n}&\lambda_{2n}\Phi_{2,2n}&\dots&\lambda_{2n}^{n-1}\Phi_{1,2n}&\lambda_{2n}^{n-1}\Phi_{2,2n}\\
\end{matrix}\right| \label{4.49}
\end{eqnarray}
\begin{eqnarray}T_{n11}=
\tiny\left|\begin{matrix}1&0&\lambda&0&\dots&\lambda^{n-1}&0&\lambda^n\\
\Phi_{1,1}&\Phi_{2,1}&\lambda_1\Phi_{1,1}&\lambda_1\Phi_{2,1}&\dots&\lambda_1^{n-1}\Phi_{1,1}&\lambda_1^{n-1}\Phi_{2,1}&\lambda_1^{n}\Phi_{1,1}\\
\Phi_{1,2}&\Phi_{2,2}&\lambda_2\Phi_{1,2}&\lambda_2\Phi_{2,2}&\dots&\lambda_2^{n-1}\Phi_{1,2}&\lambda_2^{n-1}\Phi_{2,2}&\lambda_2^{n}\Phi_{1,2}\\
\Phi_{1,3}&\Phi_{2,3}&\lambda_3\Phi_{1,3}&\lambda_3\Phi_{2,3}&\dots&\lambda_3^{n-1}\Phi_{1,3}&\lambda_3^{n-1}\Phi_{2,3}&\lambda_3^{n}\Phi_{1,3}\\
\Phi_{1,4}&\Phi_{2,4}&\lambda_4\Phi_{1,4}&\lambda_4\Phi_{2,4}&\dots&\lambda_4^{n-1}\Phi_{1,4}&\lambda_4^{n-1}\Phi_{2,4}&\lambda_4^{n}\Phi_{1,4}\\
\vdots&\vdots&\vdots&\vdots&\vdots&\vdots&\vdots&\vdots\\
\Phi_{1,2n-1}&\Phi_{2,2n-1}&\lambda_{2n-1}\Phi_{1,2n-1}&\lambda_{2n-1}\Phi_{2,2n-1}&\dots&\lambda_{2n-1}^{n-1}\Phi_{1,2n-1}&\lambda_{2n-1}^{n-1}\Phi_{2,2n-1}&\lambda_{2n-1}^{n}\Phi_{1,2n-1}\\
\Phi_{1,2n}&\Phi_{2,2n}&\lambda_{2n}\Phi_{1,2n}&\lambda_{2n}\Phi_{2,2n}&\dots&\lambda_{2n}^{n-1}\Phi_{1,2n}&\lambda_{2n}^{n-1}\Phi_{2,2n}&\lambda_{2n}^{n}\Phi_{1,2n}
\end{matrix}\right| \label{4.50}
\end{eqnarray}
\begin{eqnarray}T_{n12}=
\tiny\left|\begin{matrix}0&1&0&\lambda&\dots&0&\lambda^{n-1}&0\\
\Phi_{1,1}&\Phi_{2,1}&\lambda_1\Phi_{1,1}&\lambda_1\Phi_{2,1}&\dots&\lambda_1^{n-1}\Phi_{1,1}&\lambda_1^{n-1}\Phi_{2,1}&\lambda_1^{n}\Phi_{1,1}\\
\Phi_{1,2}&\Phi_{2,2}&\lambda_2\Phi_{1,2}&\lambda_2\Phi_{2,2}&\dots&\lambda_2^{n-1}\Phi_{1,2}&\lambda_2^{n-1}\Phi_{2,2}&\lambda_2^{n}\Phi_{1,2}\\
\Phi_{1,3}&\Phi_{2,3}&\lambda_3\Phi_{1,3}&\lambda_3\Phi_{2,3}&\dots&\lambda_3^{n-1}\Phi_{1,3}&\lambda_3^{n-1}\Phi_{2,3}&\lambda_3^{n}\Phi_{1,3}\\
\Phi_{1,4}&\Phi_{2,4}&\lambda_4\Phi_{1,4}&\lambda_4\Phi_{2,4}&\dots&\lambda_4^{n-1}\Phi_{1,4}&\lambda_4^{n-1}\Phi_{2,4}&\lambda_4^{n}\Phi_{1,4}\\
\vdots&\vdots&\vdots&\vdots&\vdots&\vdots&\vdots&\vdots\\
\Phi_{1,2n-1}&\Phi_{2,2n-1}&\lambda_{2n-1}\Phi_{1,2n-1}&\lambda_{2n-1}\Phi_{2,2n-1}&\dots&\lambda_{2n-1}^{n-1}\Phi_{1,2n-1}&\lambda_{2n-1}^{n-1}\Phi_{2,2n-1}&\lambda_{2n-1}^{n}\Phi_{1,2n-1}\\
\Phi_{1,2n}&\Phi_{2,2n}&\lambda_{2n}\Phi_{1,2n}&\lambda_{2n}\Phi_{2,2n}&\dots&\lambda_{2n}^{n-1}\Phi_{1,2n}&\lambda_{2n}^{n-1}\Phi_{2,2n}&\lambda_{2n}^{n}\Phi_{1,2n}
\end{matrix}\right| \label{4.51}
\end{eqnarray}

\begin{eqnarray}T_{n21}=
\tiny\left|\begin{matrix}1&0&\lambda&0&\dots&\lambda^{n-1}&0&0\\
\Phi_{1,1}&\Phi_{2,1}&\lambda_1\Phi_{1,1}&\lambda_1\Phi_{2,1}&\dots&\lambda_1^{n-1}\Phi_{1,1}&\lambda_1^{n-1}\Phi_{2,1}&\lambda_1^{n}\Phi_{2,1}\\
\Phi_{1,2}&\Phi_{2,2}&\lambda_2\Phi_{1,2}&\lambda_2\Phi_{2,2}&\dots&\lambda_2^{n-1}\Phi_{1,2}&\lambda_2^{n-1}\Phi_{2,2}&\lambda_2^{n}\Phi_{2,2}\\
\Phi_{1,3}&\Phi_{2,3}&\lambda_3\Phi_{1,3}&\lambda_3\Phi_{2,3}&\dots&\lambda_3^{n-1}\Phi_{1,3}&\lambda_3^{n-1}\Phi_{2,3}&\lambda_3^{n}\Phi_{2,3}\\
\Phi_{1,4}&\Phi_{2,4}&\lambda_4\Phi_{1,4}&\lambda_4\Phi_{2,4}&\dots&\lambda_4^{n-1}\Phi_{1,4}&\lambda_4^{n-1}\Phi_{2,4}&\lambda_4^{n}\Phi_{2,4}\\
\vdots&\vdots&\vdots&\vdots&\vdots&\vdots&\vdots&\vdots\\
\Phi_{1,2n-1}&\Phi_{2,2n-1}&\lambda_{2n-1}\Phi_{1,2n-1}&\lambda_{2n-1}\Phi_{2,2n-1}&\dots&\lambda_{2n-1}^{n-1}\Phi_{1,2n-1}&\lambda_{2n-1}^{n-1}\Phi_{2,2n-1}&\lambda_{2n-1}^{n}\Phi_{2,2n-1}\\
\Phi_{1,2n}&\Phi_{2,2n}&\lambda_{2n}\Phi_{1,2n}&\lambda_{2n}\Phi_{2,2n}&\dots&\lambda_{2n}^{n-1}\Phi_{1,2n}&\lambda_{2n}^{n-1}\Phi_{2,2n}&\lambda_{2n}^{n}\Phi_{2,2n}
\end{matrix}\right| \label{4.52}
\end{eqnarray}
\begin{eqnarray}T_{n22}=
\tiny\left|\begin{matrix}0&1&0&\lambda&\dots&0&\lambda^{n-1}&\lambda^n\\
\Phi_{1,1}&\Phi_{2,1}&\lambda_1\Phi_{1,1}&\lambda_1\Phi_{2,1}&\dots&\lambda_1^{n-1}\Phi_{1,1}&\lambda_1^{n-1}\Phi_{2,1}&\lambda_1^{n}\Phi_{2,1}\\
\Phi_{1,2}&\Phi_{2,2}&\lambda_2\Phi_{1,2}&\lambda_2\Phi_{2,2}&\dots&\lambda_2^{n-1}\Phi_{1,2}&\lambda_2^{n-1}\Phi_{2,2}&\lambda_2^{n}\Phi_{2,2}\\
\Phi_{1,3}&\Phi_{2,3}&\lambda_3\Phi_{1,3}&\lambda_3\Phi_{2,3}&\dots&\lambda_3^{n-1}\Phi_{1,3}&\lambda_3^{n-1}\Phi_{2,3}&\lambda_3^{n}\Phi_{2,3}\\
\Phi_{1,4}&\Phi_{2,4}&\lambda_4\Phi_{1,4}&\lambda_4\Phi_{2,4}&\dots&\lambda_4^{n-1}\Phi_{1,4}&\lambda_4^{n-1}\Phi_{2,4}&\lambda_4^{n}\Phi_{2,4}\\
\vdots&\vdots&\vdots&\vdots&\vdots&\vdots&\vdots&\vdots\\
\Phi_{1,2n-1}&\Phi_{2,2n-1}&\lambda_{2n-1}\Phi_{1,2n-1}&\lambda_{2n-1}\Phi_{2,2n-1}&\dots&\lambda_{2n-1}^{n-1}\Phi_{1,2n-1}&\lambda_{2n-1}^{n-1}\Phi_{2,2n-1}&\lambda_{2n-1}^{n}\Phi_{2,2n-1}\\
\Phi_{1,2n}&\Phi_{2,2n}&\lambda_{2n}\Phi_{1,2n}&\lambda_{2n}\Phi_{2,2n}&\dots&\lambda_{2n}^{n-1}\Phi_{1,2n}&\lambda_{2n}^{n-1}\Phi_{2,2n}&\lambda_{2n}^{n}\Phi_{2,2n}
\end{matrix}\right|. \label{4.53}
\end{eqnarray}
The corresponding $n$-fold DT is given by 
\begin{eqnarray}
A^{[n]}_{0}&=&A_0+\frac{i}{(n-1)!}\left[\sigma_3,\frac{\partial^{n-1}T_{n}}{\partial \lambda^{n-1}}\right], \label{4.54}\\
B_{1}^{[n]}&=&B_{1}-4\epsilon_{2}(T|_{\lambda=0})_{y}, \label{4.55}\\
B_{-1}^{[n]}&=&T_n|_{\lambda=-\omega} B_{-1}T_n^{-1}|_{\lambda=-\omega}. \label{4.56}
\end{eqnarray}

\section{Soliton solutions}

 Having the explicit form of the DT, we are ready to construct exact solutions of the (2+1)-dimensional HMBE. As an example, let us present the one-soliton solution. To get the one-soliton solution we take the seed solution as  
 \begin{eqnarray}
q=v=w=p=0, \quad \eta=1. \label{5.1}
\end{eqnarray}
Then the corresponding associated linear system takes the form
 \begin{eqnarray}
\psi_{1x}&=&-i\lambda \psi_1,\label{5.2}\\
\psi_{2x}&=&i\lambda \psi_2,\label{5.3}\\
\psi_{1t}&=&(2\epsilon_1\lambda+4\epsilon_2\lambda^2)\psi_{1y}+\frac{i}{\lambda+\omega}\psi_1,\label{5.4} \\
\psi_{2t}&=&(2\epsilon_1\lambda+4\epsilon_2\lambda^2)\psi_{2y}-\frac{i}{\lambda+\omega}\psi_2.\label{5.5} 
\end{eqnarray}  
This system admits the following exact solutions
\begin{eqnarray}
\psi_{1}&=&e^{-i\lambda_{1} x+i\mu_{1} y+i[(2\epsilon_1\lambda+4\epsilon_2\lambda^2)\mu_{1}+\frac{1}{\lambda_{1}+\omega}]t+\delta_1+i\delta_2},\label{5.6}\\
\psi_{2}&=&e^{i\lambda_{1} x-i\mu_{1} y-i[(2\epsilon_1\lambda+4\epsilon_2\lambda^2)\mu_{1}+\frac{1}{\lambda_{1}+\omega}]t-\delta_1-i\delta_2+i\delta_0}\label{5.7}
\end{eqnarray}
or
\begin{eqnarray}
\psi_{1}&=&e^{\theta_{1}+i\chi_{1}},\label{5.8}\\
\psi_{2}&=&e^{\theta_{2}+i\chi_{2}}.\label{5.9}
\end{eqnarray}
 Here  $\mu_{1}=\eta+i\nu$, $\lambda_{1}=\alpha+i\beta $, $\delta_i$  are real constants, 
\begin{eqnarray}
\theta_{1}&=&\beta x-\nu y-\{\eta(2\epsilon_{1}\beta+8\epsilon_{2}\alpha\beta)+\nu[2\epsilon_{1}\alpha+4\epsilon_{2}(\alpha^{2}-\beta^{2})]+\frac{\beta}{(\alpha+\omega)^{2}+\beta^{2}}\}t+\delta_{1}\label{5.10}\\
\chi_{1}&=&-\alpha x+\eta y +\{\eta(2\epsilon_{1}\alpha+4\epsilon_{2}(\alpha^{2}-\beta^{2})]-\nu[2\epsilon_{1}\beta+8\epsilon_{2}\alpha\beta]+\frac{\alpha
+\omega}{(\alpha+\omega)^{2}+\beta^{2}}\}t+\delta_{2}\label{5.11}
\end{eqnarray}
and $\theta_{2}=-\theta_{1}, \quad \chi_{2}=-\chi_{1}+\delta_{0}$.
 Then the one-soliton solution of the (2+1)-dimensional HMBE (\ref{3.1})-(\ref{3.5}) takes the form
 \begin{eqnarray}
q^{[1]}& = &-2im_{12}, \label{5.12}\\
v^{[1]}&=&4i\epsilon_{1}m_{11y}+4\epsilon_{2}(-2im_{11}m_{11y}+4im_{12}^{*}m_{12y}), \label{5.13}\\
w^{[1]}&=&-4i\epsilon_{2}m_{11y},\label{5.14}\\
\eta^{[1]}&=& \frac{|\omega+m_{11}|^{2}-|m_{12}|^{2}}{\square}, \label{5.15}\\
p^{[1]}&=& \frac{2 m_{12}(\omega+m_{11})}{\square}, \label{5.16}
\end{eqnarray}
where 
\begin{eqnarray}
m_{11}&=&\alpha+i\beta\tanh[2\theta_{1}],\label{5.17}\\
m_{12}&=&\frac{i\beta e^{2i\chi_{1}-i\delta_{0}}}{\cosh[2\theta_{1}]},\label{5.18}
\end{eqnarray}
and
\begin{eqnarray}
\square=\omega^{2}+2\alpha\omega+\alpha^{2}+\beta^{2}=(\omega+\alpha)^{2}+\beta^{2}. \label{5.19}
\end{eqnarray}
Finally let us present the one-soliton solutions for some particular cases. 

a) If $\epsilon_{1}=1, \quad \epsilon_{2}=0$ then the HMBE (\ref{3.1})-(\ref{3.5}) turn to the (2+1)-dimensional Schr\"odinger-Maxwell-Bloch equation (SMBE) of the form
 \begin{eqnarray}
iq_{t}+q_{xy}-vq-2ip&=&0, \label{5.20}\\
v_{x}+2\delta(|q|^2)_y&=&0,\label{5.21}\\
p_{x}-2i\omega p -2\eta q&=&0,\label{5.22}\\
\eta_{x}+\delta(q^{*} p +p^{*} q)&=&0,\label{5.23}
 \end{eqnarray}
 where $\delta=1$ for  our case. Its one-soliton solution has the form
 \begin{eqnarray}
q^{[1]}& = &-2im_{12}^{\prime}, \label{5.24}\\
v^{[1]}&=&4im_{11y}^{\prime}, \label{5.25}\\
\eta^{[1]}&=& \frac{|\omega+m_{11}^{\prime}|^{2}-|m_{12}^{\prime}|^{2}}{\square}, \label{5.26}\\
p^{[1]}&=& \frac{2 m_{12}^{\prime}(\omega+m_{11}^{\prime})}{\square}, \label{5.27}
\end{eqnarray}
where $m_{11}^{\prime}=m_{11}|_{\epsilon_{1}=1, \\ \epsilon_{2}=0}$ and $m_{12}^{\prime}=m_{12}|_{\epsilon_{1}=1, \\ \epsilon_{2}=0}$. This solution was obtained in \cite{R15}. 

b) Now let we put
 $\epsilon_{1}=0,  \epsilon_{2}=1$. In this case, the HMBE (\ref{3.1})-(\ref{3.5}) becomes the complex modified Korteweg-de Vriez-Maxwell-Bloch equation (cmKdVMBE) of the form 
   \begin{eqnarray}
q_{t}+q_{xxy}+ivq+(wq)_x-2p&=&0, \label{5.28}\\
v_{x}-2i\delta(q^{*}_{xy}q-q^{*}q_{xy})&=&0,\label{5.29}\\
w_{x}-2\delta(|q|^2)_y&=&0,\label{5.30}\\
p_{x}-2i\omega p -2\eta q&=&0,\label{5.31}\\
\eta_{x}+\delta(q^{*} p +p^{*} q)&=&0,\label{5.32}
 \end{eqnarray}
 where $\delta=+1$. The one-soliton solution of the (2+1)-dimensional cmKdVMBE follows from (\ref{5.12})-(\ref{5.16}) and has the form 
 \begin{eqnarray}
q^{[1]}& = &-2im_{12}^{\prime\prime}, \label{5.33}\\
v^{[1]}&=&4(-2im_{11}^{\prime\prime}m_{11y}^{\prime\prime}+4im_{12}^{*\prime\prime}m_{12y}^{\prime\prime}), \label{5.34}\\
w^{[1]}&=&-4im_{11y}^{\prime\prime},\label{5.35}\\
\eta^{[1]}&=& \frac{|\omega+m_{11}^{\prime\prime}|^{2}-|m_{12}^{\prime\prime}|^{2}}{\square}, \label{5.36}\\
p^{[1]}&=& \frac{2 m_{12}^{\prime\prime}(\omega+m_{11}^{\prime\prime})}{\square}, \label{5.37}
\end{eqnarray}
where $m_{11}^{\prime\prime}=m_{11}|_{\epsilon_{1}=0, \,  \epsilon_{2}=1}$ and $m_{12}^{\prime\prime}=m_{12}|_{\epsilon_{1}=0, \, \epsilon_{2}=1}$. 
\section{Conclusion}
 The DT is very useful to derive all kinds of solutions of integrable equations. Here the DT for the (2+1)-dimensional HMBE is constructed. In particular, the one-fold DT is presented in detail. The general formulas for the $n$-fold DT is also given.  Using the derived one-fold DT, the one-soliton  solution of  the (2+1)-dimensional HMBE is found. We note that using  the above presented DT, one can also construct the $n$-solitons, breathers,  rogue waves and  other type exact  solutions of  the (2+1)-dimensional HMBE.
 Finally we note that these results can be extend to the other nonlinear equations including integrable spin  systems \cite{royal}-\cite{R2001}.

\end{document}